\documentclass[twocolumn,showpacs,preprintnumbers,amsmath,amssymb,superscriptaddress,unsortedaddress]{revtex4}

 \usepackage{graphicx}
 \usepackage{dcolumn}
 \usepackage{bm}
 \usepackage{epsf,epsfig,latexsym}

\begin{document}
 \title{Measurement of Mutual Coulomb Dissociation in  $\sqrt{s_{NN}}=130$ GeV
 Au+Au collisions}

 \author{Mickey Chiu}\affiliation{Columbia University, New York, NY 10027}
 \author{Alexei Denisov}\affiliation{IHEP, Protvino, Russia}
 \author{Edmundo Garcia}\affiliation{U. of Maryland, College Park, MD 20742}
 \author{Judith Katzy}  \affiliation{MIT, Cambridge, MA 02139}
 \author{Andrei Makeev} \affiliation{Texas A$\&$M University, College
 Station, Texas  77843-3366}
 \author{Michael Murray} \affiliation{Texas A$\&$M University, College
 Station, Texas  77843-3366}
 \author{Sebastian White} \affiliation{Brookhaven National Laboratory, Upton, New
 York 11973}
 \noaffiliation

 \begin{abstract}
 We report on the first measurement of Mutual Coulomb Dissociation in
 heavy ion collisions.  We employ forward calorimeters to measure neutron
 multiplicity at beam rapidity. The
 cross-section for simultaneous electromagnetic breakup of Au nuclei at
 a nucleon-nucleon center of mass energy
 $\sqrt{s_{NN}}=130$ GeV is $\sigma_{MCD}=3.67\pm 0.26$ barns which is
 comparable to the
 geometrical cross section.
The ratio of the electromagnetic to the total cross section is
in good agreement with  calculations, as is the neutron multiplicity distribution.
\end{abstract}
 \maketitle

 An interesting aspect of the Relativistic Heavy Ion Collider (RHIC) is the
 high rate of $\gamma$-hadron collisions it produces.
 Photons from the highly Lorentz contracted electro-magnetic
 field produced by the heavy ions of one beam collide with  the nuclei
 of the other beam (for an overview see \cite{trautmann}).
 For example, the flux of equivalent
    photons with energies 2 GeV $\leq E_{\gamma} \leq 300$ GeV in the 
target nucleus rest-frame 
     corresponds to a $\gamma$-nucleus luminosity of $~ 10^{29} cm^{-2} \times
    s^{-1}$.
 With such a high flux, nuclear
 dissociation is highly probable.
 The calculated cross-section for Single Beam
 Dissociation of 95 barns
  limits the maximum beam lifetime at RHIC \cite{baltz2}.

 We report
 the first measurement of Mutual Coulomb Dissociation (MCD) whereby
 both beams dissociate electromagnetically. The calculation of this
 cross-section uses an extension of the Weizs\"acker-Williams
 formalism since it is dominated by 2nd order 2 photon exchange
 \cite{baltz3,igor1}. The total cross section
 $\sigma_{tot}$,
 for
 the nuclear break up of both beams,
 due to either Coulomb or strong (hadronic) interactions,
 was calculated in
 \cite{baltz3}.
There it was argued
  that uncertainties in the calculation of the two interactions
  partially cancel resulting in a theoretical error of $5\%$.
  One can exploit this to derive MCD and
 hadronic cross sections
 from the data presented here.
 The MCD process is also of interest
 because it can
 be used for luminosity monitoring at RHIC and  at
the Large Hadron Collider \cite{white2,white}.
 Finally an understanding of the photon flux generated in
 peripheral heavy ion collisions is necessary to understand the
 emerging field of $\gamma \gamma$, $\gamma$A and $\gamma$ pomeron
 interactions \cite{StarPeriph,LhcPeriph}.

The presented data are collected from the 3 experiments BRAHMS,
 PHENIX and PHOBOS  at RHIC with Au beams at
 $\sqrt{s_{NN}}$ = 130 GeV.
 MCD results in  emission of a few nucleons, dominantly neutrons,
 with small (few MeV) kinetic energy in the nucleus rest frame.
  In the laboratory frame they
 therefore have small angular spread with respect to the beam and
 close to 65 GeV energy.
 We measured neutron multiplicities close to
 the beam direction with Zero Degree Calorimeters (ZDCs) that
 are common  to all RHIC experiments. The topological
 cross-sections
 Au+Au$\rightarrow N^{Left}_{neutrons}+N^{Right}_{neutrons}$+X+Y
 are  measured and compared to calculations  \cite{baltz3,igor1}. In addition,
 Beam-Beam Counters which differ for each
  experiment
    were used to detect produced particles at large angles.

\begin{figure}[floatfix]
  \epsfig{file=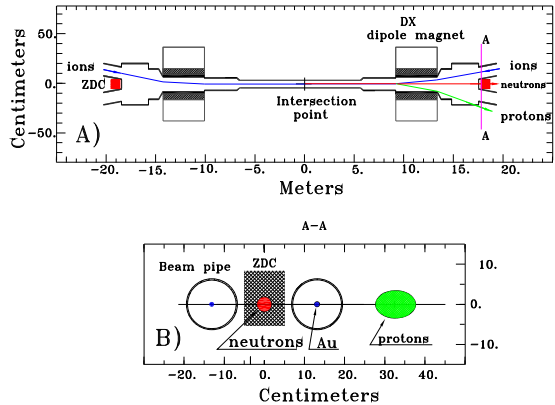,width=8.5cm}
 \caption{ \label{fig:layout}ZDC location: (a) top view
   (b) view along the beam axis. Nominal position and spread of
  neutrons, Au ions and protons at the ZDC location are shown.}
  \end{figure}

\begin{figure}[floatfix]
   \epsfig{file=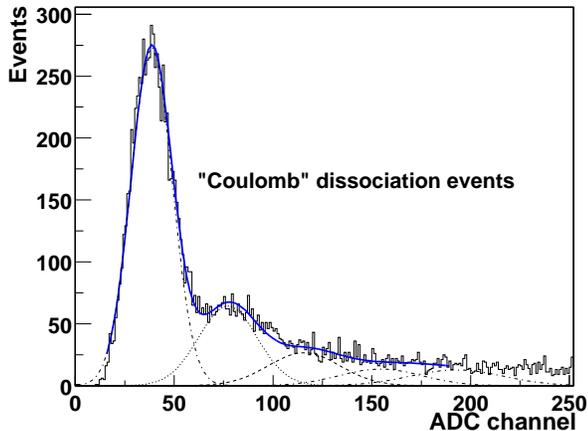,width=8.5cm}
    \caption{\label{fig:rawdist}
   Left ZDC spectrum with
   ``Coulomb" selection
 for events in which there is one neutron in the right ZDC.}
 \end{figure}

   The ZDCs  are small transverse-area hadron calorimeters with an angular
 acceptance of $|\theta|< 2$ mrad with respect to the beam axis.
 They  are located
  in the beam line behind the beam bending magnets
 at the four RHIC interaction regions and
   measure the energy of unbound neutrons in small forward cones around the   beam.
   Charged particles are deflected out of the ZDC acceptance by the beam
 bending magnet
   leading to a measurement of neutral energy with very low background.
  The layout of the ZDCs is shown in Fig.~\ref{fig:layout}
  and further details can be found in ref.~\cite{instrum}.
 The interaction trigger required an energy deposit above 12 GeV  in
 the left and right ZDCs.
 The energy scale was determined
 from fits to the one and two neutron peaks, which
 are clearly seen in Fig.~\ref{fig:rawdist}, but the  detector   response
 linearity was confirmed over the full range of observed energies.
  The energy resolution at the one neutron peak
  is approximately 21\%,
 consistent with
 test beam results \cite{instrum}. This agrees with  our calculations
 which show that there is negligible  energy deposit
 other than from neutrons in the  ZDC.

   The following discussion is based on the PHENIX analysis, while the BRAHMS
  and PHOBOS results were obtained from similar
   analyses \cite{qmtalks,brahms}. Events were selected by a ZDC
   coincidence  corresponding to a distance of $\leq 20$ cm from 
  the nominal interaction point 
   (calculated using the time-of-flight to the ZDCs).
 This
  cut is identical
 to the selection used in ref.~\cite{PXpub} and 160,601 events satisfied
  these requirements.

 The PHENIX Beam Beam Counters (BBCs)\cite{BBC} are arrays of quartz
 Cerenkov detectors which measure
 relativistic charged particles produced in cones around each beam
 ($3.05 <|\eta|< 3.85$, with $2\pi$ azimuthal coverage). There are 64
 photomultipliers (PMTs)  in each BBC arm. A coincidence of the 2 BBC arms was
 used to tag hadronic collisions for the ZDC trigger sample. 

\begin{figure}[floatfix]
 \epsfig{file=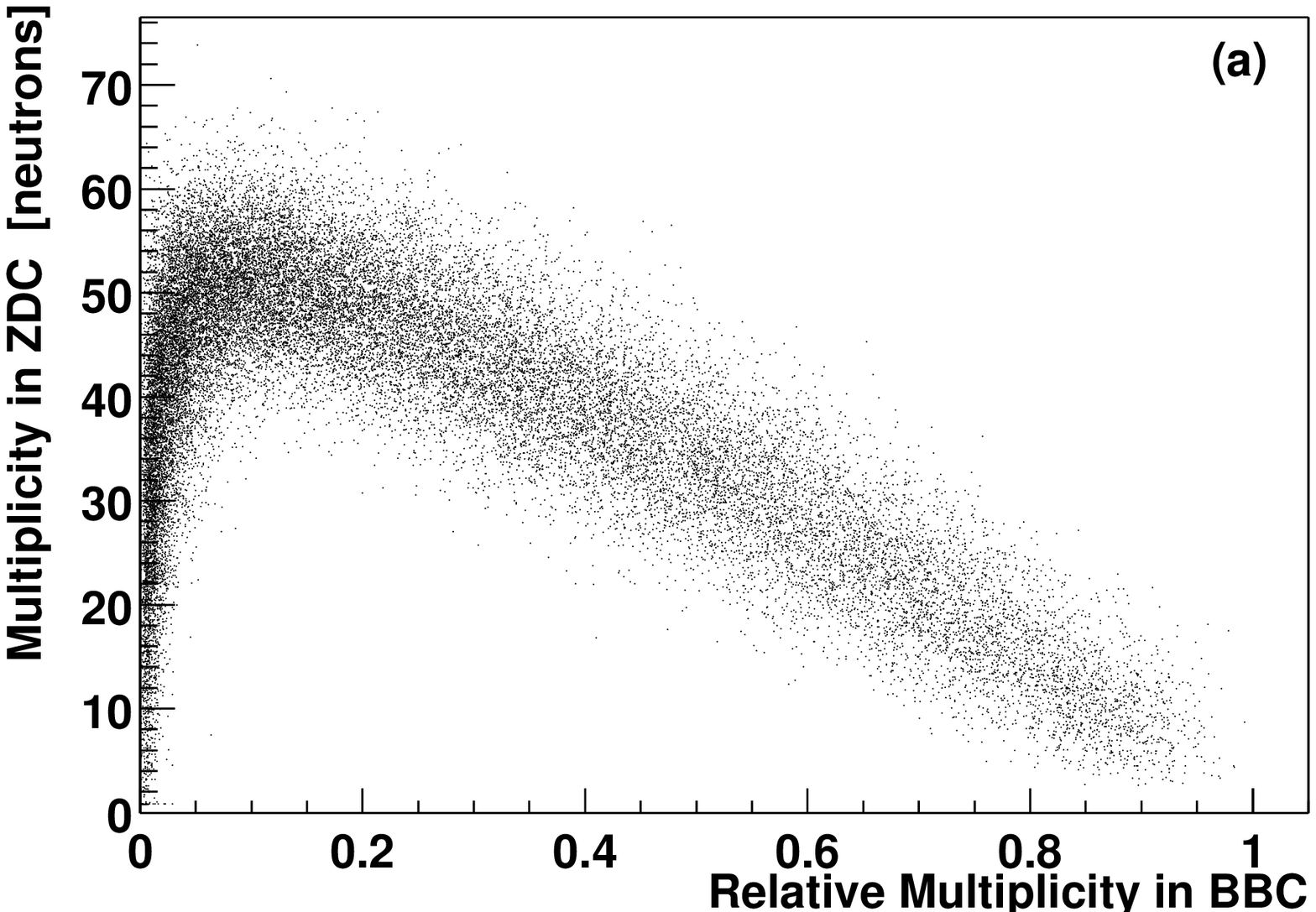,width=8.5cm}
 \epsfig{file=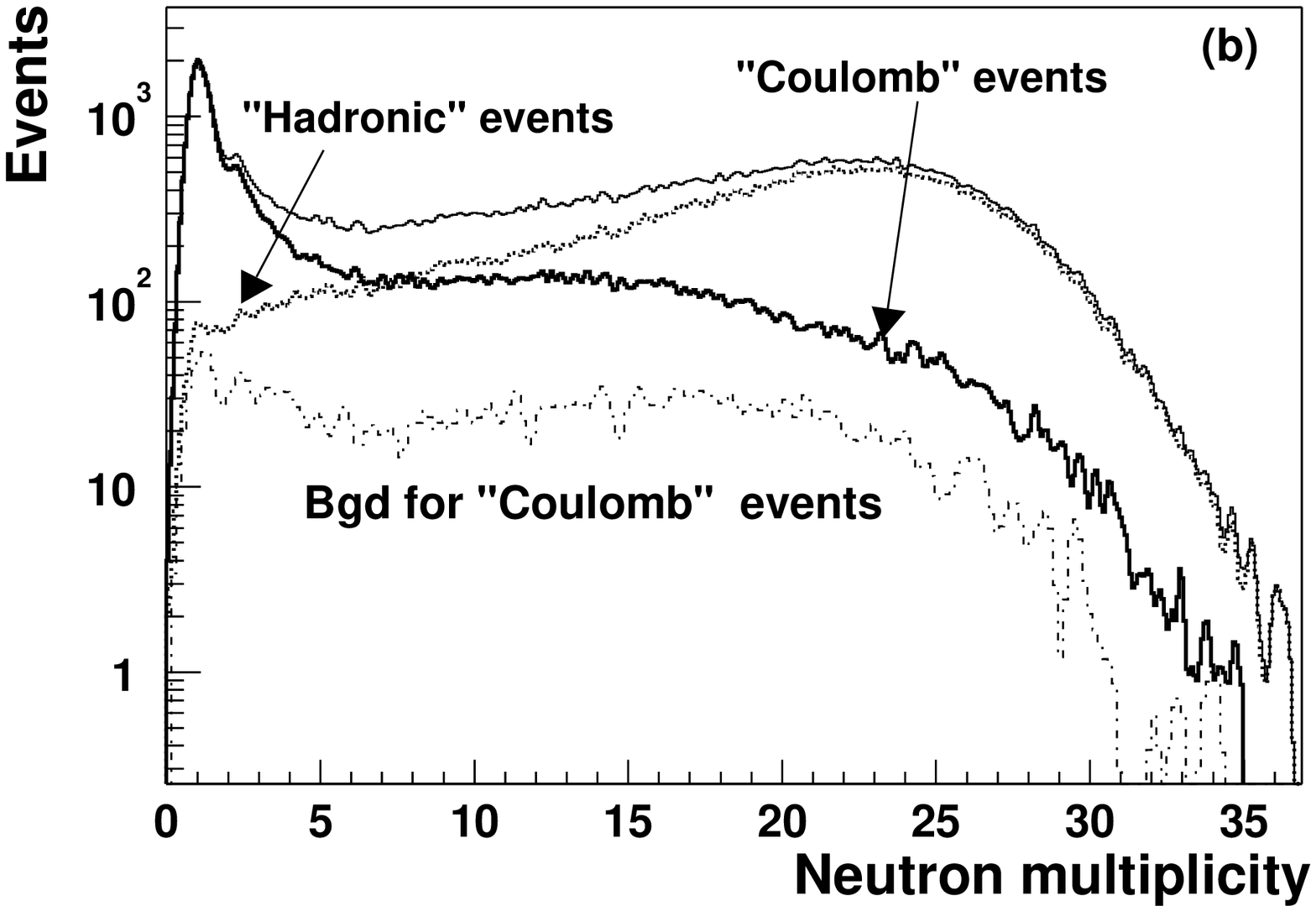,width=8.5cm,}
 \caption{\label{fig:scatter}  (a)
   Scatter plot of total ZDC neutron multiplicity (sum of 2 arms)
   (measured as energy divided by the energy of one neutron)
   versus the  BBC multiplicity.
   (b) Single arm ZDC neutron spectrum
   for the Coulomb and hadronic events.}
  \end{figure}

 In Fig.~\ref{fig:scatter}(a) we plot the neutron multiplicity measured in
 the ZDCs (i.e. total amplitude relative to the single neutron peak)
 versus the BBC multiplicity (total amplitude relative to that of a
 single hit)  \cite{footnote}.   This amplitude was then rescaled
 to set the  maximum multiplicity equal to 1.
 We define 2 classes of events according to the number of hit BBC
 photomultipliers above threshold, $n _{BBC}$
\begin{enumerate}
  \item{ ``Hadronic" events for which $n_{BBC}>1$ in each arm.}
  \item{ ``Coulomb" events with  $n_{BBC}\leq1$ in at least  one arm.}
\end{enumerate}

   The BBC trigger efficiencies for Coulomb and
 hadronic collisions were calculated from simulations of particle production
   for each class.
  We find that 99\% of Coulomb interactions satisfy
   the Coulomb cut criteria since
   only ~3\% of the events have a charged particle in either arm \cite{igor1}.
   However PHENIX calculated that
   only $92\pm2$\% of hadronic interactions passed
   the ``Hadronic" cut  defined above \cite{PXpub}.
   The other events are
   misidentified and represent a 14\% contamination of the ``Coulomb" sample.

 Diffractive hadronic interactions are not treated in our
   simulation and these are a potential background
   to our ``Coulomb''
   sample. However this is a negligible correction since, for example,
   the high energy p+W diffraction dissociation
   cross-section is 20 mb \cite{Helios}, which
   is a small fraction of the total p+W inelastic
   cross-section of 2.7b.
   Background from accidental coincidences between single beam interactions
  was calculated from the individual ZDC rates and found to be negligible.

   The ZDC multiplicity spectra are very different for the 2 classes of
  events as can be seen from Fig.~\ref{fig:scatter}(b).
 The ``Coulomb" events
 tend to have low total neutron multiplicity whereas the fraction of
   ``Hadronic" events with 1 or 2 neutrons is very small.
      Hadronic interactions which fail the ``Hadronic" cut
   are peripheral events
  corresponding to a few elementary N-N collisions so
  events with  low
 BBC multiplicity  were used to model the shape of the 14\%
 background for ``Coulomb'' events.
 The corrected ``Coulomb'' shape is identical to the ZDC
 multiplicity spectrum obtained by measuring the left ZDC distribution
   for events with one neutron   in the right one.
 This cut strongly suppresses hadronic collisions.

 The ZDC trigger efficiency correction was obtained from
 the geometrical acceptance for neutrons in the ZDC.  The angular
   distribution of neutrons about the beam direction is expected to be
   different for hadronic and Coulomb events so we discuss the two cases
   separately.
  In low  energy photoproduction experiments the neutron multiplicity increased
 with
  photon energy and
 their maximum
  $p_T$ was 120 MeV/c   \cite{gdr,veyssiere}.
 For 65GeV neutrons this corresponds to a maximum opening angle of less than 2
 mrad
 and therefore an efficiency of 100\%. However  $\sim$20\% of the
   MCD cross
   section is due to equivalent photons with energy $\geq 2$ GeV for which
   no photoproduction data are available.
   From the
   trend of the lower energy data  we would expect that these events would have a
   mean neutron multiplicity  larger than 1. Thus even if the $p_T$
 distribution broadens  at higher photon energies the   ZDC inefficiency is expected to be small since
 it is unlikely that every neutron will miss the detector.
  Therefore, no ZDC
  acceptance correction is applied to the
 measured Coulomb rates.

 The ZDC acceptance correction for hadronic collisions was 
    measured using an independent data sample which required only a
    BBC coincidence in the trigger. 
    BBC trigger
    events were scanned for interactions in which the ZDC trigger
    condition
 failed. The
    measured efficiency is $98\pm 2\%$ and this correction is applied to the
    calculated hadronic rates.
 The acceptance correction was also independently estimated assuming a
neutron $p_T$ spectrum characteristic of a Fermi
    distribution \cite{fermi}.
     The calculated
    geometrical acceptance per neutron in the ZDC detector is $75\%$ at
      $\sqrt{s_{NN}}$=130 GeV. As can be seen from Fig.~\ref{fig:scatter}(a)
       multiple neutrons are
      detected for most nuclear collisions and so
      the calculated ZDC detection
      efficiency
      per event is close to unity, consistent with the measurement.

    We corrected our
   measured ratio of ``Hadronic'' events
   to total events
   to account for the
   (experiment dependent) BBC
   trigger efficiency, $\epsilon_{BBC}$. For
    PHENIX  $\epsilon_{BBC}$  was found to be [$92\pm2(syst)$]\% ~\cite{PXpub}.
   The trigger efficiency is
   very weakly dependent on input assumptions of the
   Glauber model, such as the radius parameter. Note that it is only this
   efficiency and not the Glauber cross-section that is used in the
 current analysis.
 The BBC trigger had a
 background  contamination of [$1\pm1(syst)$]\%~\cite{PXpub}.

 Table~\ref{tab:ratio21} presents our
 measured value of the ratio of $\sigma_{geom}$ to the total ZDC cross
 section, $\sigma_{tot}$. This ratio agrees well 
  with calculations  in \cite{baltz3,igor1}.
 To compare to specific channels calculated for Mutual Coulomb
 Dissociation, we fit the measured ZDC energy spectra to neutron
 multiplicity distributions, convoluted with the energy resolution.
 We constrained the fits such
 that $\sigma_E(2n)=\sqrt{2}\times \sigma_E(1n)$, etc.
 This is justifiable from first principles. However the fitted areas can vary
 significantly if this constraint on the relative peak widths
 is removed and this variation dominates  our systematic errors.

   Figure~\ref{fig:rawdist} shows the energy spectrum obtained in one ZDC (ZDC
 left) for the ``Coulomb" event selection when the other ZDC pulse height
 is consistent with 1 neutron.
 The total number of events in Fig.~\ref{fig:rawdist}, after background
 subtraction, corresponds to the cross-section for the (1,X) topology
 in which one neutron
 is observed in the right beam direction.
 Table~\ref{tab:ratio21} lists  the sum of
 1 neutron in left
 and right topologies as  (1,X). Of course, any number of
  protons could also be emitted
   in the reaction but these are not detected.
   The decay topology (1,1) corresponds to  exactly 1 neutron
  in ZDC left and 1 neutron in ZDC right.
  Similarly the fits to specific neutron multiplicities in the ZDCs such as
  (2,1) are derived from  fits to the left ZDC spectrum, for events where we
 have cut on the pulse height in the right ZDC.
 The errors shown include both statistical and
 systematic errors in the fit procedure.
  \begin{table}[floatfix]
\renewcommand{\arraystretch}{1.28}
\begin{tabular}{cccccc}
$\mathbf{\sigma _i}$   & {\bf PHENIX} & {\bf PHOBOS}
& {\bf  BRAHMS} & {\bf  \cite{baltz3} }& {\bf   \cite{igor1}}\\ \hline
$\sigma_{tot}$                            &         -- & -- &
 -- &$10.8\pm 0.5$ & 11.2  \\
 $\sigma_{geom}$                        &  --  & -- & --
 &$7.1$ & 7.3\\
 $ \frac{\sigma_{geom}}{\sigma_{tot}}$
 & .661 $\pm$.014 & $.658 \pm .028 $ & $.68 \pm .06$ &  .67 &
 .659 \\
 $\frac{\sigma(1,X)}{ \sigma_{tot}} $
 & .117 $\pm$  .004 & .123 $\pm$ .011 & .121 $\pm$ .009 &    .125
 & .139  \\
 $\frac{\sigma(1,1)}{ \sigma(1,X)} $
 & $.345 \pm  .012 $ & $.341 \pm .015$ & $.36 \pm .02$ & $.329$
  & -- \\
  $\frac{\sigma(2,X)}{ \sigma(1,X)} $& $.345 \pm .014$
 & $.337 \pm .015 $ & $.35 \pm .03$ & -- &.327 \\
 $\frac{\sigma(1,1)}{ \sigma_{tot}} $
 & .040 $\pm$  .002 & .042 $\pm$ .003 & .044 $\pm$ .004 &    .041  $\pm$  .002
 & -  \\
\end{tabular}
\caption{\label{tab:ratio21} Ratios of cross sections for experiment and theory. The values
of $\sigma_{tot}$ and
 $\sigma_{geom}$  are in barns.}
\end{table}

 For luminosity measurements the (1,1) topology is useful because it
  has a very low  contamination from hadronic events,
 see Fig.~\ref{fig:scatter}(b), and only
 a small parameter dependence in the
 theoretical calculation \cite{baltz3}.
 Table~\ref{tab:ratio21} shows that for the ratio  $\sigma(1,1)/ \sigma_{tot}$
 the three experiments agree well with each other and the calculation.
  The error on the
 theoretical  value is dominated by the error on $\sigma_{tot}$.

\begin{figure}[floatfix]
  \epsfig{file=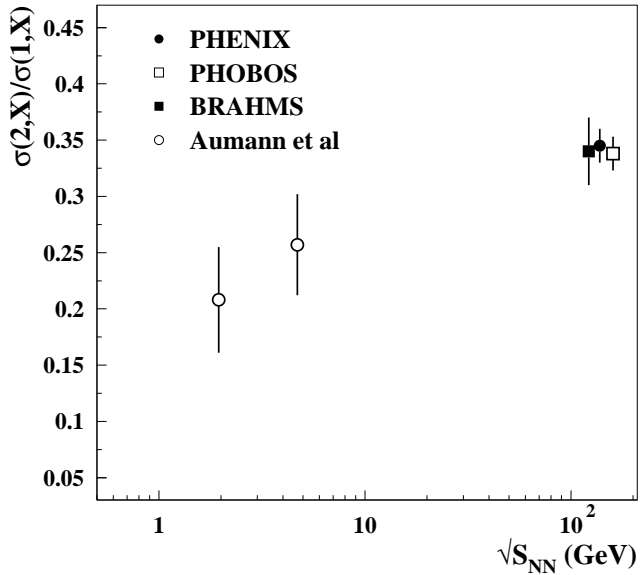,width=8.5cm}
  \caption{  \label{fig:sdep}
  Ratio of 2n to 1n cross-sections vs.  $\sqrt{s_{NN}}.$}
 \end{figure}

   In Fig.~\ref{fig:sdep} we compare our measured 2n  to 1n
 ratio to lower energy data on single beam dissociation \cite{hill}.
 The ratio tends to increase with energy and the values
  are in reasonable agreement with calculations found in    \cite{baltz3,igor1}.

   Calculations of    $\sigma_{MCD}$  suffer from the need to
   impose an arbitrary cutoff in the impact parameter  $b_0$ below
   which electromagnetic
   processes are ignored and the
   interaction is assumed to be only hadronic.
   Reducing  $b_0$
  increases $\sigma_{MCD}$  but leaves $\sigma_{tot}$  little changed.
  Also the error on
   $\sigma_{tot}$ is reduced due to the fact that
  at a given impact parameter
   the probability for either an electromagnetic or hadronic interaction
   cannot exceed one. Using this technique gives
   $\sigma_{tot} = 10.8\pm0.5$b at $\sqrt{s_{NN}}=130$ GeV  \cite {baltz3}.
   Since this result is more precise than current measurements of
    $\sigma_{geom}$ we  use it to calculate $ \sigma_{MCD}$,
 \begin{equation}
   \sigma_{MCD}=\sigma_{tot}-\sigma_{geom}=\sigma_{tot}\cdot(1-
  \sigma_{geom}/\sigma_{tot})
 \label{eq:smcd}
 \end{equation}
   and the error on $\sigma_{MCD}$ has contributions from the
   measured ratios and the
 \underline{theoretical} error on $\sigma_{tot}$.
 Using a weighted mean of the  measurements of
 $\sigma_{geom}/\sigma_{tot}$ by the 3 experiments in Eqn.~\ref{eq:smcd} gives
 $\sigma_{MCD}=3.67 \pm 0.26$ barns.

\begin{figure}[floatfix]
   \epsfig{file=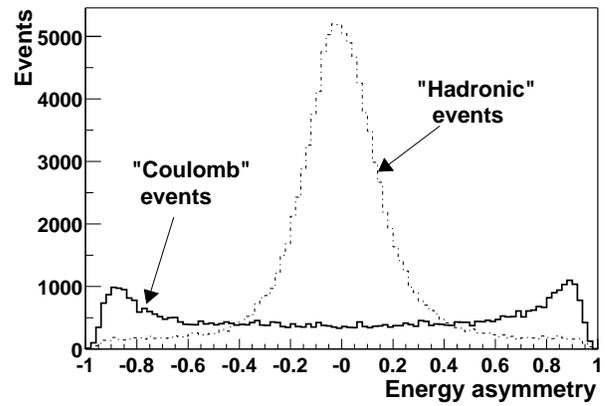,width=8.5cm}
    \caption{\label{fig:C2EMHad}
    Asymmetry spectra for Coulomb and hadronic events.}
   \end{figure}

    In Coulomb events each nucleus interacts independently with the field
   of the other.
   We therefore expect only a weak correlation between fragment
   multiplicities of the left
   and right going nuclei.  For hadronic collisions both nuclei have
   the same average number of ``wounded nucleons" and therefore the left and right
   energy in each event should be symmetric.
  We define the asymmetry
  \begin{equation}
   A(E_{Left},E_{Right})  \equiv
   \frac{E_{Left}-E_{Right}}{E_{Left}+E_{Right}}
   \label{eq:Asym}
   \end{equation}
 where $E_{Left}$ and $E_{Right}$ are the total ZDC energy signal in
   the Left and Right ZDCs
   respectively.
    Note we require
   $>98$GeV in at least 1 of the ZDCs to
   overcome the correlation inherent in the trigger.
   Figure~\ref{fig:C2EMHad} shows that hadronic events are much more symmetric
   than hadronic ones.
   We have also calculated the  correlation function
   \begin{equation}
   C_2(E_{Left},E_{Right})  \equiv
   \frac{P(E_{Left},E_{Right})}{P(E_{Left})\cdot P(E_{Right})}
   \label{eq:c2def}
   \end{equation}
  where
 $P(E_{Left},E_{Right})$ is the joint probability to have energies
  $E_{Left}$ and $E_{Right}$ in each ZDC and
  $P(E_{Left})$ is the  corresponding single probability.
  $C_2$ is flat
 and $\approx 1$ for Coulomb events but reaches  4.6 for hadronic ones.

In conclusion, we observe a class of events in Au+Au collisions at RHIC 
 where neutrons are detected at beam rapidity but few particles are observed 
 at large angles.
    The cross-section is  comparable to the geometrical  cross
 section and the ratio of the cross sections is in good agreement with earlier
 calculations of Mutual
   Coulomb Dissociation. Other features such as the neutron multiplicity
   distributions are also reasonably well reproduced.

    We thank the staff of the RHIC project (particularly A.Drees) and of
   the Collider-Accelerator
   and Physics Departments at BNL. We also thank our colleagues on
 the BRAHMS, PHENIX, PHOBOS and ZDC projects. We thank Tony Baltz and Igor
   Pshenichnov for many useful discussions.
   The US Department of Energy and the Deutsche Forschungsgemeinschaft
   supported this work.


\begin{thebibliography}{99}
\bibitem{footnote} This plot differs somewhat from
   the corresponding one in
   ref ~\cite{PXpub}, where the BBC PMTs were run at high gain for
  good timing  but  suffered some pulse height saturation.
\end{thebibliography}
\end{document}